\documentclass[12pt]{article}

\usepackage{epsfig, cite}
\usepackage{amsfonts}
\usepackage{amsopn}
\usepackage{amsmath}
\usepackage{verbatim,amsthm}

\title
{\vskip -50 pt
\begin{flushright}
\normalsize\rm NORDITA-2010-108
\end{flushright}
\vskip 20 pt
Linear space of spinor monomials and realization of the Nambu-Goldstone
 fermion in the Volkov-Akulov and Komargodski-Seiberg Lagrangians}

\author{
A. A. Zheltukhin $^{a,b,c}$\thanks{e-mail: aaz@physto.se}  \\  \\
$^a$ Kharkov Institute of Physics and Technology, \\
1, Akademicheskaya St., Kharkov, 61108, Ukraine \\  
$^b$ Fysikum, AlbaNova, Stockholm University, \\
106 91, Stockholm, 
$^c$ NORDITA,  \\
Roslagstullsbacken 23, 106 91 Stockholm, Sweden
}

\date{}

\begin{document}

\maketitle

\begin{abstract}

 The analytical algorithm previously proposed by the author 
for matching the Volkov-Akulov (VA) and Komargodski-Seiberg (KS) 
actions describing the Nambu-Goldstone (NG) fermion, is discussed.
The essence of the algotithm is explained, its consistency 
is proved and the recent results obtained 
with computer assistance are reproduced, when the proper 
Fierz rearrangements for Majorana bispinors are taken 
into account. 
We reveal a linear space of composite 
spinorial monomials $\Delta_{m}$ which are the solutions of 
the scalar constraint $(\partial^{m}\bar\psi\Delta_{m})=0$.
This space is used to clarify relations between the KS 
and VA realizations of the NG fermionic field $\psi$.

\end{abstract}

\section{Introduction}

The ideas of spontaneous supersymmetry breaking and the 
Nambu-Goldstone fermions \cite{VA} can play an important role
 in a new  physics expected from LHC. In particular, it is of
a great interest to study relevant phenomenological
 Lagrangians invariant under supersymmetry.  
Another topical problem is the relation between the  
linear and non-linear realizations of supersymmetry \cite{WB},
the study of which  resulted in new approaches 
\cite{FZ, Z, IK1, R, LR, UZ, SW, CL, CDDFG}.
Some of these achievements were taken into account and extended further 
 in the recent paper by  Komargodski and Seiberg \cite{KS},
 where the $D=4 \  {\cal N}=1$  supercurrent approach  was proposed 
to construct the low-energy effective Lagrangian. 
This approach stimulated some new proposals  
concerning the role the NG fermions and their couplings 
in the minimal standard supersymmetric model,  astroparticle 
physics and others (see e.g. \cite{ADGT}, \cite{AB}).
 
In the recent paper \cite{Zhep} we started the study  of the
direct relation between the KS and VA Lagrangians
 and developed an analytic alghorithm for proving their equivalency.
This alghorithm reduces the equivalency problem to the solution of
 three nonlinear equations \cite{Zhgen} for the three Majorana bispinors
which represent the KS fermion in terms of the VA one.
We solved these equations using the factorization property of 
these algorithmic equations. 
This shows that the use of the Majorana bispinor representation 
might be more suitable for  
 finding out a geometric principle which connects
 the KS and VA realizations of the NG fermion. 
On the contrary, working in the frame of the   2-spinor Weyl 
formalism hides deeper such a geometric origin. 
In some cases the use of the Weyl spinors greately complicates 
the  proof of the  universality  of the NG fermion action 
  in practice (see e.g. \cite{HK}, \cite{KM}, \cite{L3W}).

Here we clarify the essence of the analytical algorithm, prove its 
 consistency and show how the recent results, obtained 
with computer assistance and the Weyl spinors \cite{KT}, 
are reproduced in our approach,
 when the proper Fierz rearrangements for Majorana bispinors are taken 
into account. We reveal a linear space of composite 
spinorial monomials $\Delta_{m}$, which play an important role in the 
 redefiniton of the KS fermion via the  VA fermion $\psi$.
The spinorial monomials $\Delta_{m}$ appear as  the solution of 
the scalar constraint $(\partial^{m}\bar\psi\Delta_{m})=0$.

\section{The algorithm for matching the VA and KS actions}

Let us remind the analytical and  algorithmic procedure \cite{Zhep} for 
derivation of the relation between the VA and KS 
realizations of the NG fermions and their actions. 

The VA Lagrangian \cite{VA}, expressed in terms of the Majorana bispinors, is as  
 follows 
\begin{eqnarray} \label{VAlag}
{\cal L}_{VA}= 
 \frac{1}{a} - \frac{i}{4}\bar\psi^{,m}\gamma_{m}\psi
 - \frac{a}{32}[(\bar\psi^{,m}\gamma_{m}\psi)^2{} -
(\bar\psi^{,n}\gamma_{m}\psi)(\bar\psi^{,m}\gamma_{n}\psi)
 ]
+  \frac{a^2}{3!}\sum_{p} (-)^{p} T_{m}^{m} T_{n}^{n} T_{l}^{l}, 
\end{eqnarray} 
 where the sum  $\sum_{p}$ in (\ref{VAlag}) corresponds to the sum in 
all  the  permutations of the subindices in the products of the 
tensors $T_{n}^{m}$ defined as
\begin{eqnarray}\label{T}
 T_{m}^{n}= -\frac{i}{4}\bar\psi^{,n}
\gamma_{m}\psi,\,\,\, 
\bar\psi_{a}^{,n}:=\partial^{n}\bar\psi_{a}.
\end{eqnarray} 
The KS Lagrangian \cite{KS} expressed in the bispinor Majorana 
representation \cite{Zhep} is 
\begin{eqnarray}\label{KSoriglag} 
\begin{array}{c}
{\cal L}_{KS}=
 - f^2 -\frac{i}{2}{\bar G}^{,m}\gamma_{m}G
- \frac{1}{4f^2}[({\bar G}^{,m}G)^{2} + ({\bar G}^{,m}\gamma_{5}G)^{2}]\\
- \frac{1}{(16f^3)^2}[({\bar G}G)^{2} + ({\bar G}\gamma_{5}G)^{2}]
[(\partial^{2}({\bar G}G))^2 +((\partial^{2}({\bar G}\gamma_{5}G))^{2}],
\end{array}
\end{eqnarray}
where the  total derivative terms are omitted.
For matching the Lagrangians (\ref{KSoriglag}) and (\ref{VAlag}) 
we redefine the KS field  $g:= \sqrt{2}G$ and use arbitrariness 
in the relation between the interaction constants  $-1/f^2:=a/2$ 
resulting in  
 the  equivalent representation of the KS Lagrangian
\begin{eqnarray} \label{KSlag} 
\begin{array}{c}
{\cal L}_{KS}=
 \frac{2}{a} - \frac{i}{4}{\bar g}^{,m}\gamma_{m} g
+ \frac{a}{32}[({\bar g}^{,m}g)^{2} + ({\bar g}^{,m}\gamma_{5}g)^{2}]\\
- 2(\frac{a}{16})^3[({\bar g}g)^{2} + ({\bar g}\gamma_{5}g)^{2}]
[(\partial^{2}({\bar g}g))^2 +((\partial^{2}({\bar g}\gamma_{5}g))^{2}].
\end{array}
\end{eqnarray} 

A congruent structure of the Lagrangians $(\ref{VAlag})$ and $(\ref{KSlag})$, 
observed in \cite{Zhep}, was used there to relate the fields $g$ and $\psi$.
 It means that  the  Lagrangians contain  the fermions and 
their derivatives in the form 
of the  powers $(\partial\bar\psi\psi)^{r}$ and $(\partial{\bar g}g)^{r}$.
 The only difference consists in various distributions of the spinor indices, 
 derivatives and $\gamma$-matrices which  generate such  Lorentz covariants.
 The dimensions of these monomials are measured  
by  the degrees  $L^{-4r}$ that are inverse to the dimensions of $a^r$. 
This key information was encoded in the  general polynomial 
expansion for the Majorana bispinor $g_a$ in degrees of $\partial\bar\psi\psi$  
\begin{eqnarray}\label{redef}
g= \psi + a\chi +  a^2\chi_{2} + a^3\chi_{3},
\end{eqnarray}
where $\chi_{1}\equiv\chi$.
The maximal degree of the polynomial (\ref{redef}) is equal to three. 
Really, since the products $ a^r\chi_{r}$ have the same dimension 
as $\psi$, the Grassmannian bispinors  $\chi, \, \chi_{2} , \, \chi_{3}$ 
must have the general form 
 $\chi_{r}\sim\psi(\partial\bar\psi\psi)^{r}$ fixed by the 
above-discussed monomials. 
These monomials are nilpotent and their maximal degree $r=3$, 
because $\psi(\partial\bar\psi\psi)^{3}$ contains the maximal power 
of the  Grassmannian bispinor $\psi$ equal to four for the considered
 case $D=4 \  {\cal N}=1$.
 The substitution of (\ref{redef}) in the KS Lagrangian 
(\ref{KSlag}) and comparison of its terms with those
 in (\ref{VAlag}) with the same degree in the constant 
 $a$, yield the coupled chain of the 
three equations for $\chi, \, \chi_{2}$ and  $\chi_{3}$, which  
 was reduced  \cite{Zhgen} to the factorized form 
\begin{eqnarray}\label{factor}
(\bar\psi^{,m}\gamma_{m}\chi_{r}) = \bar\psi^{,m}{\cal P}^{(r)}_{m},
\,\,\,\,\, \,(r=1,2,3).
 \end{eqnarray} 
The polynomials ${\cal P}^{(r)}_{m}(\psi,\,\chi_{r-1},\, \chi_{r-2})$ 
(\ref{factor}), found from the previous steps of the 
algorithmic procedure, 
depend on $\psi,\,\,\chi_{r-1},\,\,\chi_{r-2}$, their 
derivatives and are considered to be known.

The presence of the derivative $\bar\psi^{,m}$ in the l.h.s and r.h.s. of 
Eqs. (\ref{factor}) was used in  \cite{Zhep}, \cite{Zhgen} 
for their simplification by 
means of cancellation of $\bar\psi^{,m}$.
These cancellations reveal representations for $\gamma_{m}\chi_{r}$ 
that 
can be associated with particular solutions of the 
inhomogeneous Eqs. (\ref{factor}).
Then the general solution of (\ref{factor}) is presented
 by the sum 
\begin{eqnarray}\label{gensol}
\gamma_{m}\chi_{r}= {\cal P}^{(r)}_{m} + \Delta^{(r)}_{m},\,\,\,\,\, (r=1,2,3)
\end{eqnarray}
of the particular solutions and the general 
solution  $\Delta^{(r)}_{m}$ of the homogeneous system
\begin{eqnarray}\label{hfactor}
(\bar\psi^{,m}\gamma_{m}\chi_{r}) =0,
\end{eqnarray} 
 formed by the linear combination of the 
bispinors $\Delta^{(r)i}_{m}$ with the complex constants $\alpha^{(r)i}$
\begin{eqnarray}\label{consts}
\Delta^{(r)}_{m}= \sum_{i}
\alpha^{(r)i}\Delta^{(r)i}_{m}.
\end{eqnarray}
 The  bispinors $(\Delta^{(r)i}_{m})_a$ in (\ref{consts})
form a linear space $\Delta^{(r)}$ of the
monomials $\sim\psi(\partial\bar\psi\psi)^{r}$, similar to the monomilas 
composing $\chi_{r}$, but satisfying the orthogonality constraints 
\begin{eqnarray}\label{constr}
(\bar\psi^{,m}\Delta^{(r)i}_{m})=0,\,\,\,\,\,\,  (r=1,2,3).
\end{eqnarray}

In general, the constraints (\ref{constr}) can be weakened 
to take into account that 
 the KS and VA Lagrangians might be equal modulo total derivatives 
\begin{eqnarray}\label{tdt}
(\bar\psi^{,m}\Delta^{(r)i}_{m}) \sim \partial^{n}\phi^{(r)i}_{n},
\,\,\,\,\,\,
\phi^{(r)i}_{n} \sim \bar\psi\psi(\partial\bar\psi\psi)^{r}, 
\end{eqnarray}
where  $\phi^{(r)i}_{n}$ are composite Lorentz vectors built of $(r+2)$ 
bispinors $\psi$ and of $r$ partial derivatives $\partial\psi$. 
Because of the nilpotency of these monomials the weakening concerns 
 the monomials with $r<3$. 
The conditions (\ref{tdt}) can  be written  in the form generalizing 
(\ref{constr})
\begin{eqnarray}\label{weakcon}
(\bar\psi^{,m}\Delta^{(r)i}_{m})=0 \,(mod\,\, total \,\,derivatives),
\end{eqnarray}
where the total derivatives in (\ref{weakcon}) are the Lorentz invariant 
monomials $\sim \partial[\bar\psi\psi(\partial\bar\psi\psi)^{r}]$ 
representable in the form of total derivatives (TD).

Taking into account the fact that the 
polynomials ${\cal P}^{(r)}_{m}$ in (\ref{factor}) are fixed by
 the algorithmic procedure,
one can conclude that the solution of Eqs. (\ref{gensol}), presented in the form
\begin{eqnarray}\label{resum1}
\gamma_{m}\chi_{r}= {\cal P}^{(r)}_{m} + \Delta^{(r)}_{m}, \\
(\bar\psi^{,m}\Delta^{(r)}_{m})=0 \,(mod\,\,TD^{(r)}) \label{resum2},
\end{eqnarray}
is reduced to the construction of the three linear 
spaces $\Delta^{(r)}$ for each of r=(1,2,3).

Below, we demonstrate the work of the algorithm by explicit 
calculation of $\chi$  from (\ref{redef}).

\section{Explicit construction of the space $\Delta^{(1)}$}

Here we build an explicit realization of the $\Delta^{(1)}$ space, 
introduced in \cite{Zhep}, and find $\chi$.

In this case the system  (\ref{factor}) is reduced to the only 
one equation 
\begin{eqnarray}\label{eqn1}
i(\bar\psi^{,m}\gamma_{m}\chi) = 
 \frac{1}{16}[({\bar\psi}^{,m}\psi)^{2} + ({\bar\psi}^{,m}\gamma_{5}\psi)^{2}] + 
\frac{1}{16}[(\bar\psi^{,m}\gamma_{m}\psi)^2{} -
(\bar\psi^{,n}\gamma_{m}\psi)(\bar\psi^{,m}\gamma_{n}\psi)]
\end{eqnarray}
which represents the equivalency conditions in the 
 first order in $a$ and  has the discussed  factorizable form. 
The cancellation of the derivative $\bar\psi^{,m}$ reduces 
  Eq. (\ref{eqn1}) to the form
\begin{eqnarray}\label{eqn2}
\gamma_{m}\chi= -\frac{i}{16}[\psi({\bar\psi}_{,m}\psi) + 
\gamma_{5}\psi({\bar\psi}_{,m}\gamma_{5}\psi)] -
 \frac{i}{16}[\gamma_{m}\psi (\bar\psi^{,n}\gamma_{n}\psi) -\gamma_{n}\psi
(\bar\psi^{,n}\gamma_{m}\psi)] + \Delta_{m},\\
(\bar\psi^{,m}\Delta_{m})=0 \,(mod\,\,TD) \label{eqn3}. 
\,\,\,\,\,\,\,\,\,\,\,\,\,\,\,\,\,\,\,\,\,\,\,\,\,\; \;\;
\,\,\,\,\,\,\,\,\,\,\,\,\,\,\,\,\,\,\,\,\,\,\,\,\,\,\,\, 
\end{eqnarray}

The system (\ref{eqn2}) consists of eight complex equations for 
two complex components of the bispinor $\chi$ and 
six complex coefficients $\alpha^{i}\equiv \alpha^{(1)i}$ 
from  (\ref{consts}) 
\begin{eqnarray}\label{consts1}
\Delta_{m}=\sum_{i}\alpha^{i}\Delta^{i}_{m},
\end{eqnarray}
 associated with the linear space of the 
monomials $\Delta^{i}_{m}\equiv \Delta^{(1)i}_{m},\, (i=1,2,..,6)$, 
which form $\Delta_{m}$
for the case $r=1$. The bispinors $\Delta^{i}_{m}$ are built
 from the monomials $\sim \psi(\partial\bar\psi\psi)$
 and may be split in two different subsets forming the linear 
space $\Delta^{(1)}$.

The first subset is presented by the  three 
monomials $\nu^{i}_{m}, (i=1,2,3)$ 
beloning to the exact solutions of (\ref{eqn3}) 
that are not accompanied 
with total derivatives in the r.h.s. of (\ref{eqn3})
\begin{eqnarray}\label{nueqn}
(\bar\psi^{,m}\nu^{i}_{m})=0.
\end{eqnarray}
The expicit form of these  monomials $\nu^{i}_{m}$ is as follows  
\begin{eqnarray}\label{nu1}
\begin{array}{c}
\nu^{1}_{m}=\varepsilon_{mnpq}\gamma^{n}[\gamma^{5}\psi(\bar\psi^{,p}\gamma^{q}\psi)
- \psi(\bar\psi^{,p}\gamma^{q}\gamma^{5}\psi)],
\\
\nu^{2}_{m}=\varepsilon_{mnpq}[\gamma^{5}\psi(\bar\psi^{,n}\Sigma^{pq}\psi)
+ \Sigma^{pq}\psi(\bar\psi^{,n}\gamma^{5}\psi)],
\\
\nu^{3}_{m}=\psi(\bar\psi^{,n}\Sigma_{nm}\psi) - \Sigma_{mn}\psi(\bar\psi^{,n}\psi).
\end{array}
\end{eqnarray}
The second subset is presented by  three solutions  $\Delta^{i}_{m}, (i=1,2,3)$ 
of Eq.(\ref{eqn3}) that are  accompanied with the   
total derivative terms in the r.h.s. of (\ref{eqn3}) having the form
\begin{eqnarray}\label{3td}
\begin{array}{c}
\Delta^{1}_{m}=\frac{1}{2}\varepsilon_{mnpq}\gamma^{n}
\partial^{p}[\psi(\bar\psi\gamma^{5}\gamma^{q}\psi)],
\\
\Delta^{2}_{m}=\frac{1}{2}\varepsilon_{mnpq}\Sigma^{pq}\partial^{n}
[\psi(\bar\psi\gamma^{5}\psi)],
\\
\Delta^{3}_{m}=\frac{1}{2}\Sigma_{mn}\partial^{n}[\psi(\bar\psi\psi].
\end{array}
\end{eqnarray}
The bispinors $\Delta^{i}_{m}(\ref{3td})$ have a remarkable property 
 to remain to be total derivatives after 
their contraction with the derivative $\bar\psi^{,m}$ 
of the bispinor $\bar\psi$
\begin{eqnarray}\label{rettdt}
\begin{array}{c}
(\bar\psi^{,m}\Delta^{1}_{m})=\frac{1}{2}\varepsilon_{mnpq}\partial^{p}[
(\bar\psi^{,m}\gamma^{n}\psi)(\bar\psi\gamma^{5}\gamma^{q}\psi)],
\\
(\bar\psi^{,m}\Delta^{2}_{m})=\frac{1}{2}\varepsilon_{mnpq}\partial^{n}[
(\bar\psi^{,m}\Sigma^{pq}\psi)(\bar\psi\gamma^{5}\gamma^{q}\psi)],
\\
(\bar\psi^{,m}\Delta^{3}_{m})=\frac{1}{2}\partial^{n}
[(\bar\psi^{,m}\Sigma_{mn}\psi)(\bar\psi\psi)].
\end{array}
\end{eqnarray}

The explicit form of the bispinors (\ref{nu1}-\ref{3td}), saturating the scalar 
constraint (\ref{eqn3}), results in the 
desired monomial representation of $\Delta_{m}$  (\ref{consts1}) 
\begin{eqnarray}\label{gentdt}
\Delta_{m}= \alpha_{1}\nu^{1}_{m}+ \alpha_{2}\nu^{2}_{m} +\alpha_{3}\nu^{3}_{m} +
 \beta_{1}\Delta^{1}_{m}+ \beta_{2}\Delta^{2}_{m} +\beta_{3}\Delta^{3}_{m}.
\end{eqnarray}
 including six arbitrary complex constants $\alpha_{i}$ and $\beta_{i}$.
The substitution of (\ref{gentdt}) in equations 
 (\ref{eqn2}) fixes position of the arbitrary 
constants $\alpha_{i},\,\beta_{i}$ in the r.h.s. of Eqs. (\ref{eqn2}) 
\begin{eqnarray}\label{eqn2'}
\begin{array}{c}
\gamma_{m}\chi= - \frac{i}{16}[\psi({\bar\psi}_{,m}\psi) + 
\gamma_{5}\psi({\bar\psi}_{,m}\gamma_{5}\psi)] -
 \frac{i}{16}[\gamma_{m}\psi (\bar\psi^{,n}\gamma_{n}\psi) -\gamma_{n}\psi
(\bar\psi^{,n}\gamma_{m}\psi)]  \\
+  \sum\alpha_{i}\nu^{i}_{m}  +  \sum\beta_{i}\Delta^{i}_{m}.
\end{array}
\end{eqnarray}

The general solution for $\chi,\, \alpha_{i}, \,\beta_{i}$ following from Eqs.
(\ref{eqn2'}) is considered below.

\section{Fierz rearrangements and general solution}

In the case of nonlinear realization 
of supersymmetry by the NG field $\psi$,  the VA Lagrangian is invariant
modulo total derivatives (see e.g. \cite{WB}). 
This points to possible presence of the monomials from 
the space $\Delta^{(1)}$ in the VA Lagrangian or in its contribution 
\begin{eqnarray}\label{VAtdt}
\zeta_{m}:=\gamma_{m}\psi (\bar\psi^{,n}\gamma_{n}\psi) -\gamma_{n}\psi
(\bar\psi^{,n}\gamma_{m}\psi)
\end{eqnarray} 
in  Eqs. (\ref{eqn2'}), denoted  by a  condenced bispinor $\zeta_{m}$.    
If the  total derivatives, discovered in $\zeta_{m}$ (\ref{VAtdt}),
 would coincide with the bispinors $\nu^{i}_{m}$ (\ref{nu1})  
or  $\Delta^{i}_{m}$ (\ref{3td}), 
they could be unified with these bispinors resulting in 
a simplification of the 
 system (\ref{eqn2'}). 
Consequently, the next step  is to verify possible presence 
of the monomials from $\Delta^{(1)}$-space  in
 the bispinor $\zeta_{m}$ (\ref{VAtdt}).

To this end consider the following Fierz rearrangement 
of $\zeta_{m}$ (\ref{VAtdt}) 
\begin{eqnarray}\label{fierz1}
\zeta_{m}=
- \frac{1}{4} \sum_{A}
\left \{\gamma_{m}\Gamma^{A}\psi(\bar\psi^{,n}\gamma_{n}\Gamma_{A}\psi)
-\Gamma^{A}\gamma_{m}\psi(\bar\psi^{,n}\Gamma_{A}\gamma_{n}\psi)\right \}.
\end{eqnarray}
The 16 Dirac matrices $\Gamma^{A}$ and their 
inverse $\Gamma_{A}=(\Gamma_{A})^{-1}$, defined as 
\begin{eqnarray}\label{basis}
\begin{array}{c}
\Gamma^{A}:=(1,\, \gamma^{m}, \, \Sigma^{mn},  \, \gamma^{5},\,
\gamma^{5} \gamma^{m}),  \\ 
\Gamma_{A}:=(\Gamma^{A})^{-1}=(1,\, -\gamma_{m}, 
\, -\Sigma_{mn},  \, -\gamma^{5},\, -\gamma^{5} \gamma_{m}), 
\end{array}
\end{eqnarray}
form the complete basis in the space of $4\times4$ matrices. 
The r.h.s. of (\ref{fierz1}) includes only the products of
$\gamma^{m}\times\Gamma^{A}$ and is
expressed via their 
(anti)commutators using  the identities 
\begin{eqnarray}\label{anti_com}
 \gamma^{m}\Gamma^{A}= \frac{1}{2}\{\gamma^{m}, \Gamma^{A}\}
+ \frac{1}{2}[\gamma^{m}, \Gamma^{A}].
\end{eqnarray}
This yields the following representation for $\zeta_{m}$ 
\begin{eqnarray}\label{fierz2}
\zeta_{m}=
- \frac{1}{8} \sum_{A}
\left \{ [\gamma_{m},\Gamma^{A}]\psi(\bar\psi^{,n}\{\gamma_{n},\Gamma_{A}\}\psi) 
-\{\Gamma^{A},\gamma_{m}\}\psi(\bar\psi^{,n}[\Gamma_{A},\gamma_{n}]\psi)\right \}
\end{eqnarray}
 including only the products of the commutators with the anticommutators. 

The desired simplification of (\ref{fierz2}) is achieved by  
substitution  in the relations
\begin{eqnarray}\label{coantico}
 \begin{array}{ll}
\{\gamma_{m}, \gamma_{n} \}= -2\eta_{mn},&
[\gamma_{n}, \Sigma_{pq}]= -2(\eta_{np}\gamma_{q}- \eta_{nq}\gamma_{p}),\\
\{\gamma_{n}, \Sigma_{pq} \}= -2\varepsilon_{npql}\gamma^{l}\gamma^{5},&
\gamma^{5}\Sigma_{mn}=-\frac{1}{2}\varepsilon_{mnpq}\Sigma^{pq},
\end{array}
\end{eqnarray}
where $\Sigma^{mn}: = \frac{1}{2}[\gamma_{m}, \gamma_{n}]$,    
 accompanied with the  use of  the explicit 
form of $\nu^{i}_{m}$ (\ref{nu1}).

As a result, we reveal the presence of the bispinors from  $\Delta^{(1)}$
in $\zeta_{m}$ (\ref{fierz2})
\begin{eqnarray}\label{finzet}
 \begin{array}{c}
\zeta_{m}
\equiv\gamma_{m}\psi (\bar\psi^{,n}\gamma_{n}\psi) -\gamma_{n}\psi
(\bar\psi^{,n}\gamma_{m}\psi)  \\
=  - [\Sigma_{mn}\psi(\bar\psi^{,n}\psi)
+ \gamma^{5}\Sigma_{mn}\psi(\bar\psi^{,n} \gamma^{5}\psi)]
+ (\nu^{1}_{m} -\frac{1}{4}\nu^{2}_{m} -  \frac{1}{2}\nu^{3}_{m})
\end{array}
\end{eqnarray}
which coincide with the bispinors $\nu^{i}_{m}$ (\ref{nu1}). 

The substitution of (\ref{finzet}) in Eqs. (\ref{eqn2'})
 transforms the latter into the equations 
\begin{eqnarray}\label{eqn2''}
\begin{array}{c}
\gamma_{m}\chi= - \frac{i}{16}[\psi({\bar\psi}_{,m}\psi) + 
\gamma^{5}\psi({\bar\psi}_{,m}\gamma^{5}\psi)] +
 \frac{i}{16}[\Sigma_{mn}\psi(\bar\psi^{,n}\psi)
+ \gamma^{5}\Sigma_{mn}\psi(\bar\psi^{,n}\gamma^{5}\psi)] 
+ \tilde\Delta_{m},
\end{array}
\end{eqnarray}  
where $\tilde\Delta_{m}$ differ from $\Delta_{m}$ (\ref{gentdt}) 
by shifts of the  coefficients $\alpha_{i}$ 
\begin{eqnarray}\label{gentdt'}
\begin{array}{c}
\tilde\Delta_{m}:= (\alpha_{1} - \frac{i}{16})\nu^{1}_{m} + 
(\alpha_{2} + \frac{i}{64})\nu^{2}_{m} +
(\alpha_{3} + \frac{i}{32})\nu^{3}_{m} 
+ \sum\beta_{i}\Delta^{i}_{m}.
\end{array}
\end{eqnarray}
The sum of the  two first terms in the r.h.s. of  (\ref{eqn2''}) 
can be combined in  a compact form 
\begin{eqnarray}\label{combin}
\begin{array}{c}
 - \frac{i}{16}[\psi({\bar\psi}_{,m}\psi) + 
\gamma^{5}\psi({\bar\psi}_{,m}\gamma^{5}\psi)] +
 \frac{i}{16}[\Sigma_{mn}\psi(\bar\psi^{,n}\psi)+\gamma^{5}\Sigma_{mn}\psi
(\bar\psi^{,n}\gamma^{5}\psi)] \\ 
=
\frac{i}{16}(-\eta_{mn}+\Sigma_{mn})
[\psi(\bar\psi^{,n}\psi) + \gamma^{5}\psi(\bar\psi^{,n}\gamma^{5}\psi)],
\end{array}
\end{eqnarray}
 simplifying  Eqs. (\ref{eqn2''}) to the  equations 
\begin{eqnarray}\label{simpl1}
\begin{array}{c}
\gamma_{m}\chi= \frac{i}{16}(-\eta_{mn}+\Sigma_{mn})
[\psi(\bar\psi^{,n}\psi)+\gamma^{5}\psi(\bar\psi^{,n}\gamma^{5}\psi)]
+\tilde\Delta_{m}.
\end{array}
\end{eqnarray}
 Contraction  of Eqs. (\ref{simpl1}) with  $\gamma^{m}$ and 
the use of the relation 
$(-\eta_{mn}+\Sigma_{mn})= \gamma_{m}\gamma_{n}$
result in the following representation of the sought-for bispinor $\chi$
 \begin{eqnarray}\label{solchi}
\chi= \frac{i}{16}\gamma_{n}[\psi(\bar\psi^{,n}\psi)
+\gamma^{5}\psi(\bar\psi^{,n}\gamma^{5}\psi)]
-\frac{1}{4}\gamma^{m}\tilde\Delta_{m}.
\end{eqnarray}

The substitution of the representation (\ref{solchi}) back in 
Eqs. (\ref{simpl1}) reveals  homogeneous equations for still 
unknown coefficients $\alpha_{i}$ 
and  $\beta_{i}$ included in $\tilde\Delta_{m}$  (\ref{gentdt'})
\begin{eqnarray}\label{coeffs}
(\eta_{mn}+ \frac{1}{3}\Sigma_{mn})\tilde\Delta^{n}=0.
\end{eqnarray}

One can observe that the rescaled 
matrix $(\eta_{mn}+ \frac{1}{3}\Sigma_{mn})$ 
from (\ref{coeffs}), acting on the vector and spinor 
indices of the bispinor $\tilde\Delta_{m}$, has the property $P^2=P$
\begin{eqnarray}\label{projec}
 P_{m}{}^{q}P_{q}{}^{n}=P_{m}{}^{n}, \,\,\,\,\,\, 
P_{m}{}^{n}:= \frac{3}{4}(\delta_{m}{}^{n} + \frac{1}{3}\Sigma_{m}{}^{n}),
\end{eqnarray}
because of the matrix relation 
 $\Sigma^{mq}\Sigma_{qn}= 3 \delta^{m}_{n} + 2\Sigma^{m}{}_{n}$. 

Then Eqs. (\ref{coeffs}) may be treated as a projection condition 
which requires a special form for spinor $\tilde\Delta_{m}$ (\ref{gentdt'})
 with its Lorentz vector index generated by $\gamma_{m}$ 
\begin{eqnarray}\label{monsol}
\tilde\Delta_{m}= \gamma_{m}\tilde\rho,
\end{eqnarray} 
where the bispinor $\tilde\rho$ belongs to 
the monomial space $\Delta^{(1)}$.  
The  representation (\ref{monsol}) is  derived with the help of the relation
$\Sigma_{mn}\gamma^{n}=-3\gamma_{m}$. 

The explicit form of $\tilde\Delta_{m}$, 
fixed  by (\ref{gentdt'}), (\ref{nu1}), (\ref{3td}), 
shows that its vector index is generated by either $\varepsilon_{mnpq}$
 or $\Sigma_{mn}$. 
However, the representation  (\ref{monsol}) requires 
to transfer this function into the matrix $\gamma_{m}$, and this requirement 
seems to be hardly realized by some choice of the constants  
$\alpha_{i}$ and $\beta_{i}$ in (\ref{gentdt'}). 
This hints that $\tilde\rho=0$, which implies that  Eqs. (\ref{coeffs}) 
have the only solution $\tilde\Delta_{m}=0$. 
It is suffucient to prove the consistency of the analytical approach.
The zero solution fixes the constants $\alpha_{i},\,\beta_{i}$ as
\begin{eqnarray}\label{zerosol}
\tilde\Delta_{m}=0 \,\,\,\, \Longrightarrow  
\,\,\,\,\,\, \alpha_{1} =\frac{i}{16},
\,\,\,\, \alpha_{2}= -\frac{i}{64}, 
\,\,\,\,\alpha_{3}=-\frac{i}{32},
\,\,\,\,\,\,   \beta_{i}=0.
\end{eqnarray}
This choice of the constants $\alpha_{i},\, \beta_{i}$  results in the 
explicit form of the bispinor  $\Delta_{m}$ (\ref{gentdt}),
introduced in \cite{Zhep} and belonging to  
the space $\Delta^{(1)}$, in terms of the monomials $\nu^{i}_{m}$ 
\begin{eqnarray}\label{gentdtf}
\Delta_{m}= \frac{i}{16}[\nu^{1}_{m} -  \frac{1}{4}\nu^{2}_{m} - 
 \frac{1}{2}\nu^{3}_{m}]. 
\end{eqnarray}

The use of (\ref{zerosol}) transforms the general 
solution (\ref{solchi}) into the reduced form
\begin{eqnarray}\label{finchi}
\chi= \frac{i}{16}\gamma_{n}[\psi(\bar\psi^{,n}\psi)
+\gamma^{5}\psi(\bar\psi^{,n}\gamma^{5}\psi)]
\end{eqnarray}
 which coincides with the solution found in \cite{KT} with use of
the Weyl 2-spinor representation.

The substitution of (\ref{finchi}) into the 
system (\ref{resum1}) and repetition of the above-considered 
steps using the equivalency equations  \cite{Zhgen}, restores the 
bispinors $\chi_2$ and $\chi_3$ in the expansion (\ref{redef}). 
All that proves the consistency of the analytical algorithm 
proposed in \cite{Zhep}.

\section{Conclusion}

We analysed in details the analytical algorithm \cite{Zhep} 
based on the factorization mechanism and asserting the
 equivalency between the KS and VA Lagrangians.  
The consistency of the algorithm was analyzed and proved 
using  the Majorana spinorial monomials $\Delta^{(r)}_{m}$. 

These monomials have the form  
  $\sim  \psi(\partial\bar\psi\psi)^{r},\, (r=1,2,3)$ 
 and satisfy the scalar constraints
 $(\bar\psi^{,m}\Delta^{(r)}_{m})=0 \,(mod \,\,
total \,\,derivatives)$. 
They  form the linear spaces $\Delta^{(r)}$ 
 constructed of the NG fermion $\psi$ and its derivatives.
The appearance of these monomials, triggered by the 
factorization mechanism,
 sheds new light on the relation between
the KS and VA realizations of the NG fermionic field 
and algebraic structures associated with it. 
It would be interesting to find out a geometrical 
principle lying in the base of this algebraic structure. 
We constructed an explicit realization of the space $\Delta^{(1)}$ 
and used it together with the Fierz rearrangements to show how 
the considered analytic approach reproduces the recent 
results obtained 
with computer assistance and using the Weyl spinors \cite{KT}.

\noindent{\bf Acknowledgments}

I would like to thank Eugeniy Ivanov, Zohar Komargodski, Dmitriy Uvarov
 for their helpful comments and Cosmas Zachos for his interesting letter 
and the paper \cite{UZ}. I acknowledge e-mail correspodence with 
Sergei Kuzenko and Simon Tyler.
I am grateful to the Department of Physics of Stockholm University 
and Nordic Institute for Theoretical Physics Nordita for kind hospitality. 
This research was supported in part by Nordita.


\begin{thebibliography}{99}



\bibitem{VA}
D.V. Volkov and V.P. Akulov, JETP Letters {\bf 16}  478 (1972);
 Phys. Lett. {\bf B 46}, 109  (1973);  Theor. Math. Phys.  {\bf 18}, 28 (1974).
\bibitem{WB}
J. Wess and J. Bagger, Supersymmetry and supergravity, 
Princeton University Press, Pinceton, 1992.
\bibitem{FZ}
S. Ferrara and B. Zumino,
Nucl. Phys. {\bf B 87} 207, (1975).
\bibitem{Z}
B. Zumino,
Nucl. Phys. {\bf B 127} 189, (1977).
\bibitem{IK1}
E.A. Ivanov and A.A. Kapustnikov,
J. Phys.  {\bf A 11} 2375, (1978); \\
J. Phys.  {\bf G 8} 167, (1982).
\bibitem{R}
M. Rocek, Phys. Rev. Lett. {\bf  41} 451, (1978). 
\bibitem{LR}
U. Lindstrom and M. Rocek, Phys. Rev. {\bf D 19} 2300, (1979).
\bibitem{UZ}
T. Uematsu and K. Zachos,
Nucl. Phys. {\bf B 201} 250, (1982).
\bibitem{SW}
S. Samuel and  J. Wess,
Nucl. Phys. {\bf B 221} 153, (1983).
\bibitem{CL}
T.E. Clark  and S.T. Love,
 Phys. Rev. {\bf  54} 5723, (1996). 
\bibitem{CDDFG}
R. Casalbyoni, S. De Curtis, D. Dominici, F. Feruglio and R. Gatto,
 Phys. Lett. {\bf B 220} 569 (1989). 
\bibitem{KS}
Z. Komargodski and N. Seiberg,
JHEP {\bf 0909} 066, (2009).
\bibitem{ADGT}
I. Antoniadis, E. Dudas, D.M. Ghilencea, P. Tziveloglou,
Non-linear MSSM, arXiv:1006.1662 [hep-ph].
\bibitem{AB}
I. Antoniadis, M. Buican,
 Goldstinos, Supercurrents and Metastable SUSY Breaking in N=2 Supersymmetric 
Gauge Theories, arXiv:1005.3012 [hep-th].
\bibitem{Zhep}
A.A. Zheltukhin,
"Cancellation of  4-derivative terms in the Volkov-Akulov action",
 Phys. Rev. {\bf D 82} 085005, (2010).
\bibitem{Zhgen}
A.A. Zheltukhin,
"On equivalence of the Komargodski-Seiberg action to the Volkov-Akulov action",
 arXiv:1009.2166.
\bibitem{HK}
T. Hatanaka and S.V. Ketov,
Phys. Lett. {\bf B 58} 265, (2004).
\bibitem{KM}
S.M. Kuzenko and S.A. McCarthy, JHEP {\bf 05}  012  (2005).
\bibitem{L3W}
H. Liu, H. Luo, M. Luo and L. Wang,
Leading order actions of Goldstino fields, arXiv:1005.0231 [hep-th].
\bibitem{KT}
S.M. Kuzenko and S.J. Tyler 
"Relating the  Komargodski-Seiberg and the Akulov-Volkov action: 
Exact nonlinear field redefinition",
arXiv:1009.3298.


\end{thebibliography}
\end{document}